\documentstyle[prb,aps,multicol]{revtex}

\begin{document}
\draft

\title{Hydrodynamics beyond local equilibrium: application to electron gas}
\author{I.~V.~Tokatly\cite{MIET} and O.~Pankratov}
\address{Lehrstuhl f\"ur Theoretische Festk\"orperphysik,\\
Universit\"at Erlangen-N\"urnberg, Staudtstr. 7/B2, 91054 Erlangen,
Germany}
\date{\today} 
\maketitle

\begin{abstract}
Generalized hydrodynamic theory, which does not rest on the requirement
of a local equilibrium, is derived in the
long-wave limit of a kinetic equation. The theory bridges the whole
frequency range between the quasistatic (Navier-Stokes) hydrodynamics
and the high frequency (Vlasov) collisionless limit. In addition to
pressure and velocity the theory includes new macroscopic
tensor variables. In a linear approximation these variables describe
an effective shear stress of a liquid and the generalized
hydrodynamics recovers the
Maxwellian theory of highly viscous fluids - the media behaving
as solids on a short time scale, but as viscous fluids on long
time intervals. It is shown that the generalized hydrodynamics can be
applied to the Landau theory of Fermi liquid. Illustrative results for
collective modes in confined systems are given, which show that nonequilibrium 
effects qualitatively change the collective dynamics in comparison with
the predictions of the heuristic Bloch's hydrodynamics. Earlier improvements
of the Bloch theory are critically reconsidered. 
\end{abstract}
 

\begin{multicols}{2} 
\section{Introduction}

The idea to apply a macroscopic
hydrodynamical description to the collective dynamics of the 
inhomogeneous many-electron
systems has been first suggested in a famous work by Bloch in 1933.
\cite{Bloch} In this paper he introduced a hydrodynamic theory of
a degenerate Fermi gas as a simplest phenomenological extension of
the Thomas-Fermi model to dynamical regime. Since application of 
nonequilibrium many-body theory to inhomogeneous systems is extremely
complex, the Bloch's hydrodynamic theory (BHT) remains
popular and is in use until now. Since 1933 the BHT has been
applied to variety of kinetic problems, 
\cite{Ritchie,Bennett,Fetter,Eguiluz,Barton,Halperin,Zaremba} albeit
due to the heuristic nature of the BHT, its relation to the kinetic theory
and the range of applicability remained unclear. 

A common sense of hydrodynamical theory presumes an existence of a local 
statistical equilibrium in every point of space - the condition which
allows to express the kinetic properties of a many-body system in
simple terms of pressure, density and temperature distributions.
\cite{Landau} This condition is,
however, rarely met in electron gas where the electron-electron 
collisions commonly play a minor role. This circumstance is ignored
in the BHT. Although the applications of the Bloch's hydrodynamical
theory
\cite{Ritchie,Bennett,Fetter,Eguiluz,Barton,Halperin,Zaremba,Harris}
have shown its usefulness, its failures are also well known, in
particular a prediction of a wrong plasmon dispersion.
\cite{Fetter,Eguiluz,Barton,Halperin,Harris,Dobson,Jackson} 

Is it possible to derive a macroscopic hydrodynamics-like theory which
would be valid
beyond the condition of a local equilibrium? What is the role of
collisions and how is the static (i.e. collision dominated) limit
as well as the
high frequency (collisionless) limit recovered? These are
the questions we address in the present paper. In the short
publication \cite{hyd1} we showed that such hydrodynamics existed.
In the paper Ref.\onlinecite{hyd1}, for the sake of simplicity 
we carried out the
derivation of the hydrodynamics equations only in one dimensional case.
We also restricted the theory to the second order in the long-wave limit
of a kinetic equation, which left out the thermal conductivity and
consequently the thermal equilibration processes.

In the present paper we give a new general derivation and extend the theory
to include the heat transport. We show that the generalized
hydrodynamics can be based on Landau theory of Fermi liquid and relate 
macroscopic parameters in hydrodynamics equations (i.e.
shear and bulk moduli), to microscopic Landau parameters. We also
present transparent analitical examples of the calculation of collective 
modes in confined systems of different dimensionality.    

Let us first consider the mathematical structure of BHT and the formal
microscopic restrictions of its applicability.
The Bloch's original idea was to use the set of hydrodynamics 
equations of an ideal charged liquid to describe the dynamics of a Fermi gas. 
Only the macroscopic
variables - electron density $n({\bf r},t)$, velocity ${\bf v}({\bf r},t)$,
pressure $P$, and electrostatic potential $\varphi ({\bf r},t),$ enter the
equations:
\begin{eqnarray}
D_{t}n &+&n\nabla {\bf v}=0 , 
\label{1.1} \\
mnD_{t}{\bf v} &+&\nabla P-en\nabla \varphi =0 , 
\label{1.2} \\
\nabla ^{2}\varphi &=&4\pi en-4\pi \rho _{ext},
\label{1.3}
\end{eqnarray}
where $\rho _{ext}$ is an external charge and 
$D_{t}=\partial _{t}+{\bf v}\nabla$ is a substantial derivative.
This set of equations (continuity equation (\ref{1.1}), Euler (\ref{1.2}) and 
Poisson (\ref{1.3})
equations) becomes complete when the equation of state is added. In the
original paper \cite{Bloch} Bloch identified $P$ with the kinetic pressure
of a degenerate Fermi gas:
\begin{equation}
P=\frac{1}{5m}(3\pi^{2})^{2/3}n^{5/3}
\label{1.4}
\end{equation} 
Further improvements of BHT addressed only the equation of state 
(inclusion of exchange,
correlation and quantum gradient corrections).\cite{Zaremba}

Sometimes BHT is viewed as an approximate extension of the density
functional theory to the dynamic regime.\cite{Zaremba} However, as
we have mentioned above, the hydrodynamics theory 
(\ref{1.1})-(\ref{1.4}) exhibits an inconsistency which has been 
reflected in many papers and textbooks.
\cite{Fetter,Eguiluz,Barton,Halperin,Harris,Dobson,Jackson}
The system of equations (\ref{1.1})-(\ref{1.4}) gives the wrong velocity
coefficient $v_{0}^{2}$ in the plasmon dispersion law 
$\omega ^{2}=\omega_{p}^{2}+v_{0}^{2}q^{2}$. Instead of the correct
result $\frac{3}{5}v_{F}^{2}$ 
($v_{F}$ is the Fermi velocity) of the linear response theory (RPA)
the  BHT leads to the value  
$v_{0}^{2}=\frac{1}{3}v_{F}^{2}$. This inconsistency does not
depend on a degree of degeneracy of an electron gas and is the
general property of the hydrodynamics theory which is based on 
the equilibrium equation
of state. For example, at arbitrary degeneracy the result of hydrodynamics
is $v_{0}^{2}=v_{s}^{2}$, where $v_{s}$ is the
velocity of sound, whereas in the linear response theory
$v_{0}^{2}$ equals to the mean square of the particle velocity 
$<v_{p}^{2}>$.\cite{Jackson}

It 
has been realized \cite{Harris} that this discrepancy originates from the
assumption of a local equilibrium, which underlies the common 
hydrodynamic theory.\cite{Landau} 
The assumption allows to reduce the kinetic equation for the
distribution function $f_{{\bf p}}({\bf r},t)$
\begin{equation}
\partial _{t}f_{{\bf p}}+\frac{{\bf p}}{m}\nabla f_{{\bf p}}+
e\nabla \varphi\frac{\partial f_{{\bf p}}}{\partial {\bf p}}=
I_{{\bf p}}\left[ f_{{\bf p}}\right]   
\label{1.5}
\end{equation}
to a set of equations for macroscopic variables 
$n({\bf r},t)$,${\bf v}({\bf r},t)$
and, in general case, temperature $T({\bf r},t)$. The requirement of the
local equilibrium is fulfilled
if the characteristic time of the process $t_{pr} \sim 1/\omega $ is much
longer than the inverse collision frequency $1/\nu_c $ and the typical length 
of the inhomogeneity $L$ is greater than the mean free path $l\sim
u/\nu_c $ ($u$ is the average particle velocity)\cite{Klim} 
\begin{equation}
\omega /\nu_c \ll 1,\qquad u/L\nu_c \ll 1.  
\label{1.6}
\end{equation}
In a zero order with respect to parameters (\ref{1.6}) Eq. (\ref{1.5})
reduces to $I_{{\bf p}}\left[ f_{{\bf p}}\right] $ $=0$, which means that 
the distribution function $f_{{\bf p}}$ and, consequently, the equation of
state has a locally equilibrium form. 
As a result, the equations for the first three moments of the distribution
function i.e. density $n$, current ${\bf j}=n{\bf v}$ and stress tensor 
$P_{ij}=\delta_{ij}P$ (which is diagonal in the local equilibrium) constitute
the closed set of hydrodynamics equations for an ideal liquid.

It is clear that from the microscopic point of view BHT cannot 
be a consistent theory
since it extends the collision dominated hydrodynamics 
($\nu_{c}\rightarrow \infty$) to the electron gas where the 
collisionless (Vlasov) limit ($\nu_{c}\rightarrow 0$) is most common.
Due to the high frequency of the plasma waves at least the first of
inequalities (\ref{1.6}) is strongly violated and the tensor structure of 
$P_{ij}$ as well as that of higher moments becomes important. In a degenerate
case these tensors describe deformation of the Fermi sphere - the effect
completely ignored in the BHT. It has been shown recently \cite{Conti} 
that within a linear response theory this effect leads to an effective
shear modulus of a liquid.  
 
In this paper we present a generalized hydrodynamics which remains valid
far beyond the local equilibrium, when both conditions (\ref{1.6})
are strongly violated. The theory is restricted to
the long-wave limit of the kinetic equation and requires
inclusion of the tensors of higher moments. In Sec. II we derive a
general hierarchy of equations for the moments of the distribution function
and introduce a regular procedure of truncation of the infinite chain
of equations, which does not require the assumption of a local
equilibrium.
We also discuss the physical meaning of restrictions which make possible
the truncation.
In Sec. III we construct a hydrodynamic theory of the second order
in the long wavelength expansion and show a relation of the generalized
hydrodynamics to the theory of elasticity and to the theory of highly
viscous fluids. Sec. IV is devoted to the
generalized hydrodynamics of a charged Fermi-liquid on the basis of
Landau theory. We find the relationship between macroscopic
hydrodynamical parameters and microscopic Fermi-liquid parameters. 
Sec. V contains examples of application of
the generalized hydrodynamics to collective oscillations
of confined systems. The purpose of this section is purely
illustrative. To show clearly the nonequilibrium effects we select an
analytically solvable model of a charged Fermi liquid
in a harmonic potential of different dimensionality. We show that
deviations from the local equilibrium strongly shift the excitation
frequencies and substantially change the structure of collective modes.  
In Sec. VI we extend the theory to the fourth order
in the long wavelength expansion. It is shown that in the low
frequency limit the fourth-order theory transforms into the set of
equations of a classical hydrodynamics, which includes viscosity
and a heat transport contribution. In Sec. VII we summarize our
results.   

\section{The hierarchy of equations for the moments}

The kinetic equation (\ref{1.5}) can be transformed in an infinite chain of
equations for the moments of the distribution function.
\cite{Klim} The zeroth and the first moment are the
particle density $n({\bf r},t)$ and the current density ${\bf j}({\bf r},t)$:
\begin{equation}
n({\bf r},t) =\sum_{{\bf p}}f_{\bf p}({\bf r},t), \qquad
{\bf j}({\bf r},t) =\sum_{{\bf p}}\frac{\bf p}{m}f_{\bf p}({\bf r},t)
\label{2.1} 
\end{equation}
The velocity field can be defined in a usual way as ${\bf v}={\bf j}/n$.
In general, the moment of the $k$th order is a tensor of $k$th rank
\begin{equation}
{\bf M}^{(k)}=M^{(k)}_{i_{1}...i_{k}} =\frac{1}{m^{k-1}}\sum_{{\bf p}}
p_{i_{1}}...p_{i_{k}}f_{\bf p}, \quad k>1
\label{2.2} 
\end{equation}
It is a common practice to separate the macroscopic motion of the liquid from
the relative motion of particles via transformation to the
Lagrange comoving frame. In this frame the macroscopic velocity
is equal to zero and the particle density $n({\bf r},t)$
is the same as in the laboratory frame. The
transformed distribution function $f^{L}_{\bf p}({\bf r},t)$ is 
related to the distribution function in the laboratory frame
$f_{\bf p}({\bf r},t)$ as follows
$$
f^{L}_{\bf p}=f_{{\bf p}+m{\bf v}}
$$
and apparently has the properties
\begin{eqnarray}
\sum_{{\bf p}}f^{L}_{\bf p}({\bf r},t)&=&\sum_{{\bf p}}f_{\bf p}({\bf r},t)
=n({\bf r},t),
\label{2.3a}\\
\sum_{{\bf p}}\frac{\bf p}{m}f^{L}_{\bf p}({\bf r},t)&=& 
\sum_{{\bf p}}\frac{\bf p}{m}f_{\bf p}({\bf r},t)-
{\bf v}({\bf r},t)n({\bf r},t)=0.
\label{2.3b}
\end{eqnarray}
Eqs. (\ref{2.3a}) and (\ref{2.3b}) can be considered as a definition of
the comoving frame.

The higher moments of the transformed distribution function are defined
similar to (\ref{2.2}):
\begin{equation}
{\bf L}^{(k)}=L^{(k)}_{i_{1}...i_{k}} =\frac{1}{m^{k-1}}\sum_{{\bf p}}
p_{i_{1}}...p_{i_{k}}f^{L}_{\bf p}, \quad k>1
\label{2.3} 
\end{equation}  
Equations for moments $L^{(k)}_{i_{1}...i_{k}}$ can be easily
derived from the kinetic equation for the distribution function
$f^{L}_{\bf p}$:
\begin{eqnarray} \nonumber 
D_{t}f^{L}_{\bf p}&+&\frac{{\bf p}}{m}\nabla f^{L}_{{\bf p}} -
({\bf p}\nabla){\bf v}\frac{\partial f^{L}_{{\bf p}}}{\partial {\bf p}}\\
&-& \left(mD_{t}{\bf v}-e\nabla \varphi\right)
\frac{\partial f^{L}_{{\bf p}}}{\partial {\bf p}}=
I^{L}_{{\bf p}}\left[ f^{L}_{{\bf p}}\right],
\label{2.4}
\end{eqnarray}
where $D_{t}=\partial _{t}+{\bf v}\nabla$ is a substantial derivative.
The zeroth and the first moments of this equation give respectively 
the continuity equation and the equation for the velocity ${\bf v}$:
\begin{eqnarray}
D_{t}n &+&n\nabla _{i}v_{i}=0  
\label{2.5} \\
mnD_{t}v_{i} &+&\nabla _{j}L^{(2)}_{ij}-en\nabla _{i}\varphi =0  
\label{2.6} 
\end{eqnarray} 
In the derivation of Eqs. (\ref{2.5})-(\ref{2.6}) the constraints due to
conservation of particles, momentum and energy in a collision
process  
\begin{equation}
\sum_{{\bf p}}I_{{\bf p}}=0,\quad \sum_{{\bf p}}p_{i}I_{{\bf p}}=0,\quad
\sum_{{\bf p}}{\bf p}^{2}I_{{\bf p}}=0,  
\label{2.7}
\end{equation}
have been used. The second moment $L^{(2)}_{ij}$ is a stress tensor,
which is related to the pressure $P$ of an electron gas as 
$TrL^{(2)}_{ij}=3P$. This
tensor can be decomposed into a scalar and a traceless parts
\begin{equation}
L^{(2)}_{ij}=P\delta _{ij}+\pi _{ij},\qquad Tr\pi _{ij}=0.  
\label{2.8}
\end{equation}
The equation of motion (\ref{2.6}) allows to eliminate the electrostatic
potential $\varphi$ from the kinetic equation (\ref{2.4})
\begin{equation}
D_{t}f^{L}_{\bf p} + \frac{p_{i}}{m}\nabla_{i} f^{L}_{{\bf p}} -
\left(p_{j}\nabla_{j}v_{i} - \frac{\nabla_{j}L^{(2)}_{ij}}{n}\right)
\frac{\partial f^{L}_{{\bf p}}}{\partial p_{i}}=
I^{L}_{{\bf p}}
\label{2.9}
\end{equation}
and, consequently, from all equations for higher moments. 
The term in brackets in Eq. (\ref{2.9}) has a clear
physical meaning. This is a force which acts on a particle
in the Lagrange frame. 

Calculation of the $k$th moment  of the kinetic equation (\ref{2.9})
leads to the equation for the $k$th moment of the distribution function
$L^{(k)}_{i_{1}...i_{k}}$. The $k$th moment of the first term in the
left hand side of Eq. (\ref{2.9}) gives the substantial derivative
of the $k$th moment of the distribution function ${\bf L}^{(k)}$. 
Similarly the divergency of the $k+1$th moment ${\bf L}^{(k+1)}$
comes from the second term. Contributions from the last two
terms on the left have more complicated
tensor structure. For example, 
the $k$th moment of the first term in the brackets in Eq. (\ref{2.9}) is
\end{multicols}
\widetext
\noindent
\raisebox{12pt}[0in][0.0in]
    {$\rule{3.4in}{0.4pt}\rule{0.4pt}{6pt}$\hspace{3.6in}}
\vspace{-0.15in}
\begin{eqnarray}\nonumber
\frac{1}{m^{k-1}}\sum_{{\bf p}}p_{i_{1}}...p_{i_{k}}p_{j}
\frac{\partial f^{L}_{{\bf p}}}{\partial p_{i}}\nabla_{j}v_{i}
&=& 
-\frac{1}{m^{k-1}}\sum_{{\bf p}}f^{L}_{{\bf p}}
\frac{\partial}{\partial p_{i}}p_{i_{1}}...p_{i_{k}}p_{j}\nabla_{j}v_{i}\\
&=& 
\left[ \delta_{ii_{1}}L^{(k)}_{i_{2}...i_{k}j}+
 \delta_{ii_{2}}L^{(k)}_{i_{1}i_{3}...i_{k}j}+...+
 \delta_{ij}L^{(k)}_{i_{1}...i_{k}}\right]\nabla_{j}v_{i}
\label{2.9a}
\end{eqnarray}
\vspace{-0.25in}
\raisebox{-12pt}[0in][0.0in]
    {\hspace{3.64in}$\rule{0.4pt}{6pt}\rule[6pt]{3.4in}{0.4pt}$}
\begin{multicols}{2}
The contribution from the last term in the left-hand side of Eq. (\ref{2.9})
has a similar structure but contains the $k-1$th moment ${\bf L}^{(k-1)}$
instead of ${\bf L}^{(k)}$ in Eq. (\ref{2.9a}). Since the first moment of the 
distribution function $f^{L}_{{\bf p}}$
equals to zero (see Eq.(\ref{2.3b})) this contribution vanishes in 
equation for the second moment. Hence, in the case of $k=2$ one has
the following equation for the moments:
\begin{eqnarray} \nonumber
D_{t}L^{(2)}_{ij} &+&L^{(2)}_{ij}\nabla _{k}v_{k}+
L^{(2)}_{ik}\nabla _{k}v_{j} + L^{(2)}_{kj}\nabla_{k}v_{i}  \\
&+&\nabla _{k}L^{(3)}_{ijk}=I^{(2)}_{ij},  
\label{2.10}
\end{eqnarray}
where $I^{(2)}_{ij}$ is the second moment of the collision integral.

By the definition (\ref{2.3}), all moments $L^{(k)}_{i_{1}...i_{k}}$ are
symmetric with respect to transmutation of all indexes. All equations 
for moments must preserve this symmetry. To explicitly incorporate this 
condition it is convenient to introduce special notations for two possible
symmetric products of a symmetric $k$-th rank tensor $T^{(k)}_{i_{1}...i_{k}}$
and a vector $a_{j}$. The first one is the dot product, which decreases
the rank of a tensor from $k$ to $k-1$ and corresponds to the contraction of
the vector index $j$ with one of tensor indexes
\begin{equation}
{\bf T}^{(k)}\cdot {\bf a}\equiv 
T^{(k)}_{i_{1}...i_{k-1}j}a_{j}.
\label{2.11}
\end{equation} 
We also define a symmetric direct product of ${\bf T}^{(k)}$ and ${\bf a}$ 
which gives a tensor of the rank $k+1$ and is equal to the sum over all
transmutations of the vector index $j$ and tensor indexes $i_{1}...i_{k}$
\begin{eqnarray} \nonumber
\{{\bf T}^{(k)}\otimes {\bf a} \}_{S}&\equiv& 
T^{(k)}_{i_{1}...i_{k}}a_{j} + T^{(k)}_{j i_{2}...i_{k}}a_{i_{1}}\\
&+& T^{(k)}_{i_{1}j...i_{k}}a_{i_{2}}+...+T^{(k)}_{i_{1}...i_{k-1}j}a_{i_{k}}.
\label{2.12}
\end{eqnarray} 
These notations allow us to rewrite the moments of different terms
of the kinetic equation (\ref{2.9}) in a compact way. For example,
the $k$th moment of the first term in the round brackets in Eq. (\ref{2.9}),
which is given by Eq. (\ref{2.9a}) above, can be expressed as
\begin{eqnarray}\nonumber
\frac{1}{m^{k-1}}\sum_{{\bf p}}&p_{i_{1}}&...p_{i_{k}}p_{j}
\frac{\partial f^{L}_{{\bf p}}}{\partial p_{i}}\nabla_{j}v_{i}
\\\nonumber
&=&{\bf L}^{(k)}(\nabla \cdot {\bf v}) +
\{({\bf L}^{(k)}\cdot \nabla)\otimes {\bf v}\}_{S}.
\end{eqnarray}
Similarly, the moment of the second term in the brackets in Eq. (\ref{2.9})
is
$$
\frac{1}{m^{k-1}}\sum_{{\bf p}}p_{i_{1}}...p_{i_{k}}
\frac{\partial f^{L}_{{\bf p}}}{\partial p_{i}}\nabla_{j}L^{(2)}_{ij}
=\frac{1}{m}\{{\bf L}^{(k-1)}\otimes(\nabla \cdot {\bf L}^{(2)})\}_{S}
$$

Finally the system of equations for the moments takes the form
\begin{eqnarray}
D_{t}n &+&n\nabla \cdot {\bf v}=0 , 
\label{2.13} \\
mnD_{t}{\bf v} &+&\nabla \cdot {\bf L}^{(2)}-en\nabla \varphi =0 , 
\label{2.14} \\ \nonumber
D_{t}{\bf L}^{(2)} &+& {\bf L}^{(2)}(\nabla \cdot {\bf v}) +
\{({\bf L}^{(2)}\cdot \nabla)\otimes {\bf v}\}_{S}\\
&+& \nabla \cdot {\bf L}^{(3)} = {\bf I}^{(2)}
\label{2.15} \\ \nonumber
D_{t}{\bf L}^{(k)} &+& {\bf L}^{(k)}(\nabla \cdot {\bf v}) +
\{({\bf L}^{(k)}\cdot \nabla)\otimes {\bf v}\}_{S}
+ \nabla \cdot {\bf L}^{(k+1)}\\ &-&
\frac{1}{mn}\{{\bf L}^{(k-1)}\otimes(\nabla \cdot {\bf L}^{(2)})\}_{S}  
= {\bf I}^{(k)},
\label{2.16}
\end{eqnarray}
where the last equation (\ref{2.16}) is valid for $k>2$.

In a collision dominated case when conditions (\ref{1.6}) are
fulfilled, the high moments 
are small and the infinite chain of equations (\ref{2.13})-(\ref{2.16}) 
can be truncated 
(Chapman-Enskog or Grad methods, see Ref.\ \onlinecite{Klim}). As 
the traceless part $\pi _{ij}$ of the stress tensor (\ref{2.8})
and the third moment tensor $L^{(3)}_{ijk}$ vanish for the locally 
equilibrium distribution
function, they remain small under conditions (\ref{1.6}). In this
case tensors $\pi _{ij}$ and $L^{(3)}_{ijk}$ describe, respectively, 
viscosity and thermal conductivity in a collision dominated liquid.
However, if conditions  (\ref{1.6}) are violated, $\pi _{ij}$ and 
$L^{(3)}_{ijk}$ as
well as the higher-order moments are not small. Yet the infinite chain of
equations can be decoupled if all physical quantities are slow varying
functions of ${\bf r}$. It is seen from Eq. (\ref{2.10}) that the third moment 
$L^{(3)}_{ijk}$ enters only under a spatial derivative, hence its 
contribution is
proportional to $1/L$. This remains true for all higher moments in 
higher-order equations (\ref{2.16}). 
Thus the truncation of the chain can be guaranteed by
the smallness of the gradients of the moments instead of the smallness of
the moments itself. The dimensionless parameter of this expansion is 
\begin{equation}
\gamma \sim \frac{u}{Lmax\{\omega ,\nu_c \}}\ll 1.  
\label{2.17}
\end{equation}
It is clear that both inequalities (\ref{1.6}) can be violated while the
condition (\ref{2.17}) is fulfilled. For collisionless nondegenerate
plasma the possibility to decouple the chain of equations for moments
based on the condition similar to Eq. (\ref{2.17}) was first pointed
out in Ref.\ \onlinecite{Bernstein}, where it was referred to as ``the
low temperature approximation''.

To justify the consistency of this truncation procedure we estimate the
order of magnitude of the terms in Eq. (\ref{2.16}).  
According to Eq. (\ref{2.13}), the spatial derivative of the velocity field
${\bf v}$ is of
the order of the inverse characteristic time $1/t_{pr} \sim \omega $ hence the
first three terms in Eq. (\ref{2.16}) are proportional to $\omega$. The 
right-hand side of this equation is evidently proportional to the collision
frequency $\nu_c $. The last two terms in the left-hand side 
contain only spatial derivatives and are of the order of $u/L$. Thus the
contribution of ${\bf L}^{(k)}$ in the equation for ${\bf L}^{(k-1)}$ 
contains an additional
smallness $\gamma $ (\ref{2.17}), which means that the correction 
originating from ${\bf L}^{(k)}$ to the equation for the density (\ref{2.13})
is of the order $\gamma ^{k}$. Therefore to obtain a theory valid up to 
$\gamma ^{k}$ one should keep $k+1$ equations for ${\bf L}^{(q)}$ 
($0\le q\le k$)
with the contribution of the $k+1$-th moment and the spatial derivative of
${\bf L}^{(2)}$ being omitted in the last equation. The resulting theory 
is a generalization of hydrodynamics which is valid far from the equilibrium.

It should be mentioned that, in principle, the described truncation
procedure may not provide the closed system of equations because
the moments of the collision integral are, in general, the functionals of
the distribution function. However, for a high frequency process
($\omega/\nu_{c} \gg 1$) the contribution from the collision term disappears
and we obtain the closed set of equations of the 
"collisionless hydrodynamics" 
which corresponds to the Vlasov limit of the kinetic equation.
The collision terms become eventually important on the long time
scale, when the system approaches the equilibrium. In this limit the moments
of the collision integral are linear functions of the corresponding
moments of the distribution function.\cite{Klim} Hence under the
condition ($\omega/\nu_{c} \ll 1$) they can be presented in a close
form. Matching these two 
limits gives the hydrodynamical theory which correctly describes
the high- and the low-frequency limit and is approximately valid for
all values of $\omega/\nu_{c}$.

There is an obvious relation of the generalized hydrodynamics to the
standard linear response theory. In the high-frequency region the
collisionless hydrodynamics is a regular expansion of Vlasov
equation in terms of parameter (\ref{2.17}). Consequently
in the case of a weak perturbation the linearized hydrodynamics
equations should give the same results as the linearized Vlasov
equation. For example, the response function obtained from the linearized
system (\ref{2.13})-(\ref{2.16}) which is truncated at $k$th equation should
coincide with the microscopic response function up to 
$(u/L\omega)^{k} \sim (uq/\omega)^{k}$ (where $q \sim 1/L$ is a 
wave vector of a perturbation). However the hydrodynamics in its
general form goes beyond the linear response theory (RPA) since it is
valid for nonlinear regime. 

In this section we introduced parameter $\gamma$, which
governs the decoupling procedure. The smallness of this
parameter restricts the region of applicability of the 
generalized hydrodynamics. The condition  (\ref{2.17}) physically means 
that a particle passes the characteristic length scale $L$ in a 
time much longer than the typical time of the process $\sim 1/\omega$
or a collisional time $\sim 1/\nu_{c}$. In this case there are no
particles which move in resonance with a collective motion and may
provide an energy exchange between
single-particle and collective excitation. In other words, the contribution
of the Landau damping to the evolution of the system is small. This
is the physical restriction of the applicability of the generalized 
hydrodynamics which is (as any hydrodynamics) the theory of a 
collective motion.

\section{The second-order theory. Hydrodynamics of a degenerate Fermi gas}

The theory of the second order with respect to parameter $\gamma$
(\ref{2.17}) corresponds to neglect the third moment 
${\bf L}^{(3)}$ in Eq. (\ref{2.15}). Let us decompose the stress
tensor into the scalar and the traceless  parts (\ref{2.8}) and write 
the system of equations in terms of the fields $P$
and $\bbox {\pi}$:
\begin{eqnarray}
&D_{t}&n + n\nabla \cdot {\bf v}=0 , 
\label{3.1} \\
&m&D_{t}{\bf v} + \nabla P+\nabla \cdot \bbox {\pi}-en\nabla \varphi =0 , 
\label{3.2} \\ 
&D_{t}&P+\frac{5}{3}P\nabla\cdot {\bf v}+
\frac{2}{3}(\bbox {\pi} \cdot \nabla)\cdot {\bf v}=0, 
\label{3.3}\\ \nonumber
&D_{t}&\bbox {\pi} + \bbox {\pi}(\nabla \cdot {\bf v}) +
\{(\bbox {\pi}\cdot \nabla)\otimes {\bf v}\}_{S} - 
{\bf 1}\frac{2}{3}(\bbox {\pi} \cdot \nabla)\cdot {\bf v}\\
&+&P\left( \{\nabla \otimes {\bf v}\}_{S}-
{\bf 1}\frac{2}{3}\nabla\cdot {\bf v} \right)  = {\bf I}^{(2)},
\label{3.4}
\end{eqnarray}
where ${\bf 1}$ is the unit tensor. Eqs. (\ref{3.3}) and (\ref{3.4})
are the trace and the traceless part of Eq. (\ref{2.15}) respectively. 
Due to the conservation of energy in a collision process the second
moment of the collision integral ${\bf I}^{(2)}$ is a traceless
tensor. Consequently it contributes only to Eq. (\ref{3.4}).
Eqs. (\ref{3.1})-(\ref{3.4}) are the hydrodynamics equations
which contain two scalars $n$ and $P$, one vector ${\bf v}$
and one traceless tensor $\bbox {\pi}$. For
a charged liquid one has to add the Poisson equation (\ref{1.3}) for the
scalar potential $\varphi$.

Eigenmodes of a charge liquid (plasma waves) lie in a high frequency
region where the collisionless limit is reached. In Sec. IV 
(see also Ref.\ \onlinecite{hyd1}) it is shown that the system of equation
(\ref{3.1})-(\ref{3.4}) leads
to the correct plasmon dispersion up to the second order of  
parameter $\gamma$. This is a reflection of the
fact that the linearized Eqs. (\ref{3.1})-(\ref{3.4}) correspond to
a correct linear response theory up to the $\gamma^{2}$.

In the low frequency limit an expression for collision integral
is needed. To demonstrate how a long-time scale behavior is
recovered, we take $I_{\bf p}$ in a Krook-Bhatnager-Gross (KBG) 
approximation (see, i.e. Ref.\ \onlinecite{Rukh}) 
\begin{equation}
I_{{\bf p}}[f_{{\bf p}}]=
-\nu_c \left( f_{{\bf p}}({\bf r},t)-f_{{\bf p}}^{F}({\bf r},t)\right) ,  
\label{3.5}
\end{equation}
where $f_{{\bf p}}^{F}$ is a local equilibrium Fermi function with
position-dependent velocity, chemical potential $\mu ({\bf r},t)$ and 
temperature $T({\bf r},t)$. 
By the definition the function $f_{{\bf p}}^{F}$ is chosen to
give the same values of velocity ${\bf v}({\bf r},t)$, density
$n({\bf r},t)$ and pressure $P({\bf r},t)$, as the exact
distribution function $f_{{\bf p}}({\bf r},t)$:
\begin{eqnarray}
n({\bf r},t) &=&\sum_{{\bf p}}f_{\bf p}=\sum_{{\bf p}}f_{\bf p}^{F},
\label{3.a}\\
{\bf j}({\bf r},t) &=&\sum_{{\bf p}}\frac{\bf p}{m}f_{\bf p}=
                    \sum_{{\bf p}}\frac{\bf p}{m}f_{\bf p}^{F},
\label{3.b}\\
P({\bf r},t) &=&\sum_{{\bf p}}\frac{{\bf p}^{2}}{3m}f_{{\bf p}+m{\bf v}}=
              \sum_{{\bf p}}\frac{{\bf p}^{2}}{3m}f_{{\bf p}+m{\bf v}}^{F}.
\label{3.c}
\end{eqnarray}
Eqs. (\ref{3.a}-\ref{3.c}) guarantee that the collision integral 
(\ref{3.5}) satisfies the general properties (\ref{2.7}).
 
The second moment of $I_{{\bf p}}$ (\ref{3.5}) is equal to
\begin{equation}
{\bf I}^{(2)}=-\nu_c \left({\bf L}^{(2)} -{\bf 1}P_{F}(n,T)\right) ,  
\label{3.6}
\end{equation}
where $P_{F}(n,T)$ is the pressure of a Fermi gas with the distribution
function $f_{{\bf p}}^{F}$. By definition (\ref{3.c}), this pressure equals 
to the exact
pressure $P_{F}(n,T)=P\equiv \frac{1}{3}Tr{\bf L}^{(2)}$, hence the expression 
in the brackets in (\ref{3.6}) equals to  the traceless tensor $\bbox {\pi}$. 
Consequently, Eq. (\ref{3.6}) for ${\bf I}^{(2)}$ is equivalent to 
two equations
\begin{eqnarray}
&{\bf I}^{(2)}& =-\nu_c \bbox {\pi},  
\label{3.7} \\
&P& = P_{F}(n,T).  
\label{3.8}
\end{eqnarray}
Since $P_{F}(n,T)$ is an equilibrium pressure for given $n$ and $T$, 
the equation (\ref{3.8}) may seem to be related to the assumption of 
the local equilibrium. In fact, it is not. 
Eq. (\ref{3.8}) simply introduces the new independent scalar 
variable $T({\bf r},t)$ instead of $P({\bf r},t)$. The system can still
be in an arbitrary (up to $\gamma^{2}$) nonequilibrium state. In the
second order theory such state is uniquely described
by two scalar, one vector and one traceless tensor functions. The
traceless second rank tensor $\bbox {\pi}$ is responsible for deviations
from the local equilibrium. In a degenerate Fermi
system this tensor describes deviations of the shape of the Fermi surface 
from sphere.   

The system of equations (\ref{3.1})-(\ref{3.4}), with the collision
term (\ref{3.7}) and $P$ from Eq. (\ref{3.8}) 
transforms into the common hydrodynamics theory in the limit 
$\omega/\nu_c \ll 1$.
Indeed, as we already mentioned above the spatial derivative of ${\bf v}$
has the order of magnitude of $\omega$. Hence, the first four terms in the
left hand side of (\ref{3.4}) are proportional to $\omega \bbox {\pi}$,
the last term in the left hand side is proportional to $\omega P$ and
the right hand side is equal to $\nu_{c} \bbox {\pi}$. Consequently,
in a zero order of $\omega /\nu_c $ Eq. (\ref{3.4}) leads to 
$\pi _{ij}=0$, and Eqs.(\ref{3.1})-(\ref{3.3})  become identical to 
hydrodynamics equations for an ideal liquid:
\begin{eqnarray}
D_{t}n &+&n\nabla {\bf v}=0 , 
\label{3.9} \\
mnD_{t}{\bf v} &+&\nabla P-en\nabla \varphi =0 , 
\label{3.10}\\
D_{t}P&+&\frac{5}{3}P\nabla\cdot {\bf v}=0. 
\label{3.11} 
\end{eqnarray}
Here Eq. (\ref{3.10}) is the Euler equation and Eq. (\ref{3.11}) is 
the equation 
for the conservation of energy. For a degenerate Fermi gas this set of
equations exactly corresponds to the BHT. We emphasize again, that it
is valid only in an extremely collision dominated regime.

In the first order of $\omega/\nu_c \ll 1$ one has to neglect the
first four terms in the left hand side of Eq. (\ref{3.4}) in
comparison with the last term which is proportional to the pressure $P$. 
Thus, the first order solution of Eq. (\ref{3.4}) takes the form of the
viscosity tensor 
\begin{equation}
\pi _{ij}=-\frac{P}{\nu_c }\left( \nabla _{i}v_{j}+
\nabla _{j}v_{i}-\frac{2}{3}\delta _{ij}\nabla _{k}v_{k}\right) ,  
\label{3.12}
\end{equation}
with the viscosity coefficient $\eta =P/\nu_c $. In
this case Eqs. (\ref{3.2}) and (\ref{3.3}) are equivalent to 
Navier-Stokes equation and the equation for energy conservation in 
a viscous liquid respectively.
In Eq. (\ref{3.12}) we have turned back to common tensor notations to make the 
relation to the ordinary viscous hydrodynamics more transparent.

The formulas (\ref{3.7}), (\ref{3.8}) were obtained using the KBG 
collision integral (\ref{3.6}). They are, however, more general than the 
KBG approximation itself. As it has been mentioned at the end of the
previous section, the reason is that in the low frequency range,
where collisions are important, the second moment
$I^{(2)}_{ij}$ is always a linear function of $\pi_{ij}$.\cite{Klim} The
coefficient $\nu_c $ can thus be considered as a phenomenological parameter,
which is related to viscosity $\eta$ as $\nu_{c} = P/\eta$.

Eqs. (\ref{3.1})-(\ref{3.4}) with ${\bf I}^{(2)}$ from (\ref{3.7}) and $P$
from (\ref{3.8}) constitute the closed set of equations of generalized
hydrodynamics in the second order of $\gamma$. This set of equations
gives a correct description of the high-frequency collisionless regime.
It also leads to the classical hydrodynamics of a viscous fluid in a
low-frequency limit. Consequently it should be valid, with a reasonable 
accuracy, for intermediate regime.

However, the second order theory has an inherent inconsistency. 
Though in a low frequency limit the viscous term is recovered correctly,
the thermal conductivity
contribution $\nabla^{2}T$ (Ref.\ \onlinecite{Landau}) is still missing in the
energy conservation equation. Without this contribution the correct static
limit $\nabla T({\bf r})=0$ cannot be recovered. In fact, the term 
$\nabla^{2}T$
corresponds to correction of the fourth order of $\gamma$ in the
continuity equation. Physically this means an absence of a dissipative
flow of energy with an accuracy up to $\gamma^{2}$. To include the 
thermal conductivity one has to
consider the third and the fourth moments and neglect the fifth moment
(see Sec. VI bellow). 
 
The theory of the second order in $\gamma$ is obviously
applicable for a description of dynamics of a dense degenerate Fermi
gas. In this case the condition $T/E_{F}\ll 1$ ($E_{F}$ is the
local Fermi energy) is always satisfied and it is unnecessary to take
into account 
the third and the fourth moments. At $T=0$ Eq. (\ref{3.8}) reduces to 
$P=\frac{1}{5m}(3\pi^{2})^{2/3}n^{5/3}$ and Eq. (\ref{3.4}) transforms 
into
\begin{equation}
(\bbox {\pi} \cdot \nabla)\cdot {\bf v}=0.
\label{3.13}
\end{equation}
The equation for pressure $P$ and the condition (\ref{3.13}) 
together with the system of differential equations  
\begin{eqnarray}
&D_{t}&n + n\nabla \cdot {\bf v}=0 , 
\label{3.14} \\
&m&D_{t}{\bf v} + \nabla P+\nabla \cdot \bbox {\pi}-en\nabla \varphi =0 , 
\label{3.15} \\  \nonumber
&D_{t}&\bbox {\pi} + \bbox {\pi}(\nabla \cdot {\bf v}) +
\{(\bbox {\pi}\cdot \nabla)\otimes {\bf v}\}_{S} - 
{\bf 1}\frac{2}{3}(\bbox {\pi} \cdot \nabla)\cdot {\bf v}\\
&+&P\left( \{\nabla \otimes {\bf v}\}_{S}-
{\bf 1}\frac{2}{3}\nabla\cdot {\bf v} \right)  = -\nu_{c}\bbox {\pi},
\label{3.16}
\end{eqnarray}
provide a complete system of equations of the generalized hydrodynamics
for a degenerate Fermi gas. 

Most applications of the Bloch's hydrodynamics concerned just the
case of the degenerate Fermi gas. The set of equations
(\ref{3.13})-(\ref{3.16}) of generalized hydrodynamics should replace
the BHT, which is incorrect except the collision dominated
limit or the case of spatially homogeneous system.

There is an interesting relation of the linearized version of the
generalized hydrodynamics in a high-frequency (collisionless) limit
to the theory of elasticity. In the limit $\omega/\nu_{c}\gg 1$
linearization of the system (\ref{3.1})-(\ref{3.4}) gives
\begin{eqnarray}
\partial _{t}\delta n &+&n_{0}\nabla_{k}v_{k}=0  
\label{3.17} \\
mn_{0}\partial_{t}v_{i} &+&\nabla_{i}\delta P+
\nabla_{j}\pi_{ij}-en_{0}\nabla_{i}\varphi =0  
\label{3.18} \\
\partial_{t}\delta P &+&\frac{5}{3}P_{0}\nabla_{k}v_{k}=0  
\label{3.19} \\
\partial_{t}\pi_{ij} &+&P_{0}\left( \nabla_{i}v_{j}+
\nabla_{j}v_{i}-\frac{2}{3}\delta_{ij}\nabla_{k}v_{k}\right) =0,  
\label{3.20}
\end{eqnarray}
where $\delta n$ and $\delta P$ are deviations from the equilibrium density 
$n_{0}$ and pressure $P_{0}$. Let us introduce a
displacement vector ${\bf u}({\bf r},t)$ as $\delta n =-n_0\nabla{\bf u}$.
The continuity equation (\ref{3.17}) gives a usual relation between
velocity and displacement $\partial_t{\bf u}={\bf v}$. Introducing
a stress tensor $\sigma_{ij}=-\delta P \delta_{ij}-\pi_{ij}$ one
can rewrite Eq. (\ref{3.18}) as
\begin{equation}
mn_{0}\partial^2_{t}u_{i}-\nabla_{j}\sigma_{ij}-en_{0}\nabla_{i}\varphi=0.
\label{3.21}
\end{equation}
The relationship of the tensor $\sigma_{ij}$ to the displacement $u_i$ 
follows from Eqs. (\ref{3.19}, \ref{3.20}) and takes the same form as 
in the elasticity theory \cite{Landau1}
\begin{equation}
\sigma_{ij}= K\nabla_k u_k\delta_{ij}+\mu\left( \nabla_{i}u_{j}+
\nabla_{j}u_{i}-\frac{2}{3}\delta_{ij}\nabla_{k}u_{k}\right),
\label{3.22}
\end{equation} 
where the bulk modulus $K$ and the shear modulus $\mu$ of an electron gas are
$K=\frac{5}{3}P_0$ and $\mu = P_0$. Physically the bulk modulus is 
responsible for the increase of the energy which is caused by the 
local change of the occupied
volume in a momentum space, whereas the shear stress describes the 
deviation of the shape of this volume from a sphere. In the next section
we derive the hydrodynamics for a Fermi liquid and find the correlation
contribution to bulk and shear modulus. The elastic description of Fermi 
systems has been recently discussed in Ref.\ \onlinecite{Conti} within 
the linear response theory.

Inclusion of collisions violates the exact 
correspondence to the elasticity theory. In this case the shear stress
tensor $\sigma_{ij}-\frac{1}{3}\delta_{ij}\sigma_{kk}=-\pi_{ij}$ should be 
determined from the equation
\begin{equation} 
\partial_{t}\pi_{ij} +\mu\partial_t\left( \nabla_{i}u_{j}+
\nabla_{j}u_{i}-\frac{2}{3}\delta_{ij}\nabla_{k}u_{k}\right) =-\nu_c\pi_{ij},  
\label{3.23}
\end{equation}
which follows from Eqs. (\ref{3.20}) and (\ref{3.7}).
Eq. (\ref{3.23}) exactly coincides with the equation for the shear
stress tensor in the phenomenological theory of {\it highly viscous fluids}
by Maxwell.\cite{Landau1} These fluids (for instance, glycerin or resin) 
behave as solids at short intervals of time, 
but as viscous liquids on a large time scale (see 
Ref.\ \onlinecite{Landau1}). The generalized hydrodynamics thus
provide a surprising but clear analogy between an electron gas and
Maxwellian highly viscous fluids.

\section{The generalized hydrodynamics of a Fermi liquid}

In this section we develop the hydrodynamic description of a Fermi
liquid based on the Landau theory. Since the Landau theory is valid
only for small deviations of the quasiparticle distribution function
$n_{{\bf p}}$ from the Fermi function, only linearized version of the 
generalized
hydrodynamics can be constructed. Although such theory is formally
equivalent to the linear response approach, the hydrodynamical
formulation is generally more efficient for spatially inhomogeneous
problems. In addition, the derivation of the  generalized
hydrodynamics of a Fermi liquid is of a general interest since it 
transparently shows how correlation effects contribute  to the stress 
tensor.

We start from a kinetic equation for a charged Fermi liquid 
\cite{Pines}
\begin{equation}
\partial _{t}n_{{\bf p}}+
\frac{\partial \varepsilon_{\bf p}}{\partial {\bf p}}\nabla n_{{\bf p}}-
\nabla \varepsilon_{\bf p}\frac{\partial n_{{\bf p}}}{\partial {\bf p}}+
e\nabla \varphi \frac{\partial n_{{\bf p}}}{\partial {\bf p}}=
I_{{\bf p}},   
\label{4.1}
\end{equation}
where $\varepsilon_{\bf p}$ is the local quasiparticle energy, which
depends on the distribution function and, consequently, is a function
of spatial coordinates. The gradient of this function gives an additional
force which has its origin in the correlation effects.  

Following the derivation of the hydrodynamic equations for a Fermi gas 
(Sec. II)
we separate the macroscopic and relative motion of a liquid by
transformation to the comoving frame. The kinetic equation for
the distribution function $n_{\bf p}^{L}$ in this frame looks similar
to Eq. (\ref{2.4}), but with an additional correlation force:
\begin{eqnarray} \nonumber 
D_{t}n^{L}_{\bf p}&+&
\frac{\partial \varepsilon_{\bf p}^{L}}
{\partial {\bf p}}\nabla n^{L}_{\bf p} -
({\bf p}\nabla){\bf v}\frac{\partial n^{L}_{\bf p}}{\partial {\bf p}}\\
&-& \left(mD_{t}{\bf v}-e\nabla \varphi +\nabla \varepsilon_{\bf p}^{L}\right)
\frac{\partial n^{L}_{\bf p}}{\partial {\bf p}}=
I^{L}_{\bf p}
\label{4.2}
\end{eqnarray}
According to the Landau assumption, the quasiparticle energy is a
linear functional of the deviation of the distribution function from
the Fermi step function:
\begin{eqnarray}
\varepsilon_{\bf p}^{L}&=&\varepsilon_{\bf p}^{(0)}+\sum_{\bf p'}f_{\bf p'p}
\delta n_{\bf p'}^{L},
\label{4.3}\\
n_{\bf p}^{L}&=& n_{\bf p}^{(0)} + \delta n_{\bf p}^{L}, \qquad
n_{\bf p}^{(0)} = \theta (p - p_{F}),
\label{4.4}
\end{eqnarray}
where $\varepsilon_{\bf p}^{(0)}$ is the quasiparticle energy in the
undisturbed system and $f_{\bf p'p}$ is the quasiparticle-quasiparticle
interaction.
Linearization of the kinetic equation (\ref{4.2}) leads to the
Landau-Silin equation,\cite{Pines} written in a comoving frame
\begin{eqnarray} \nonumber 
&\partial_{t}&\delta n^{L}_{\bf p} +
\frac{p_{i}}{m^{*}}\nabla_{i}\delta n^{L}_{\bf p} -
p_{j}\nabla_{j}v_{i}\frac{\partial n^{(0)}_{\bf p}}{\partial p_{i}}\\
&-& \left(m\partial_{t}v_{i}-e\nabla_{i} \varphi +
\sum_{\bf p'}f_{\bf p'p}\nabla_{i}\delta n_{\bf p'}^{L}\right)
\frac{\partial n^{(0)}_{\bf p}}{\partial p_{i}}=
I^{L}_{\bf p}.
\label{4.5}
\end{eqnarray}
The quasiparticle mass $m^{*}$, which enters equation (\ref{4.5}), is
related to the bare fermion mass $m$ as follows \cite{Pines}
\begin{equation}
\left(\frac{\partial^{2}}{\partial p^{2}}\varepsilon_{\bf p}^{(0)}\right)^{-1}
 \equiv 
m^{*}= m\left(1 +\frac{1}{3}F_{1} \right).
\label{4.6}
\end{equation}
where $F_{1}$ is the first of Landau Fermi-liquid parameters:
\begin{equation}
F_{l} = \frac{m^{*}p_{F}}{\pi^{2}}\int \frac{d\Omega'}{4\pi} f_{\bf p'p}
P_{l}(\cos \theta), 
\label{4.7}
\end{equation}
where $\theta$ is the angle between ${\bf p}$ and ${\bf p'}$, 
$d\Omega'$ is an element of a solid angle around ${\bf p'}$ and 
$P_{l}(\cos \theta)$ are Legendre polynomials.

Zeroth and first moments of Eq. (\ref{4.5}) correspond to the continuity
equation and the equation for velocity of a Fermi liquid:
\begin{eqnarray}
\partial_{t}\delta n &+&n_{0}\nabla _{i}v_{i}=0  
\label{4.8} \\
mn_{0}\partial_{t}v_{i} &+&\nabla _{j}\delta P_{ij}-
en_{0}\nabla _{i}\varphi =0  
\label{4.9} 
\end{eqnarray}
In Eq. (\ref{4.9}) we introduced the stress tensor of a Fermi liquid 
\begin{equation}
\delta P_{ij}= \sum_{\bf p}\frac{p_{i}p_{j}}{m^{*}}\delta n_{\bf p}^{L} -
\sum_{\bf p,p'}f_{\bf p'p}
\frac{\partial n^{(0)}_{\bf p}}{\partial p_{i}}p_{j}\delta n_{\bf p'}^{L}
\label{4.10}
\end{equation} 
The first term in Eq. (\ref{4.10}) 
corresponds to the kinetic stress tensor of quasiparticles, whereas
the second term describes the interaction of qusiparticles and is
responsible for a correlation contribution to the stress.

According to the results of the previous section, the second order 
approximation (up to $\gamma^{2}$) gives a consistent hydrodynamical
theory of a degenerate Fermi system. To make the system
of equations (\ref{4.8})-(\ref{4.9}) complete one needs only an
equation for the second moment or the stress tensor.
To simplify the derivation of this equation we separate
the stress tensor $\delta P_{ij}$ into the scalar $\delta P$ and the
traceless $\pi_{ij}$ parts
\begin{equation}
\delta P_{ij}= \delta P\delta_{ij} + \pi_{ij}
\label{4.11}
\end{equation} 
First we rewrite (\ref{4.10}) as follows
\begin{eqnarray}
\delta P_{ij}&=& \frac{1}{m^{*}}\sum_{\bf p}T_{ij}({\bf p})\delta n_{\bf p}^{L}
\label{4.12} \\
T_{ij}({\bf p})&=& p_{i}p_{j} + 
\frac{m^{*}p_{F}}{\pi^{2}}\int \frac{d\Omega'}{4\pi} f_{\bf p'p}p'_{i}p'_{j}
\label{4.13} 
\end{eqnarray}
In the Landau theory all momentums reside on the Fermi surface.
By symmetry tensor $T_{ij}({\bf p})$ is uniquely representable
in the form
\begin{equation}
T_{ij}({\bf p})= \alpha p^{2}\delta_{ij} + \beta (p_{i}p_{j}-
\frac{1}{3} p^{2}\delta_{ij}),
\label{4.14}
\end{equation}  
where the constants $\alpha$ and $\beta$ are determined by the following two
equations
\begin{eqnarray}
\frac{1}{p_{F}^{2}}T_{ii}= 3\alpha = 1 &+& 
\frac{m^{*}p_{F}}{\pi^{2}}\int \frac{d\Omega'}{4\pi} f_{\bf p'p},
\label{4.15} \\ \nonumber
\frac{1}{p_{F}^{4}}p_{i}T_{ij}p_{j}= \alpha + \frac{2}{3}\beta &=&\\
1 &+&
\frac{m^{*}p_{F}}{\pi^{2}}\int \frac{d\Omega'}{4\pi} f_{\bf p'p}
\frac{({\bf pp'})^{2}}{p_{F}^{4}}.
\label{4.16} 
\end{eqnarray}
Solutions to these equations are
\begin{equation}
\alpha = 1 + F_{0}, \qquad \beta = 1 + \frac{1}{5}F_{2}
\label{4.17}
\end{equation}  
Substituting $\alpha$ and $\beta$ (\ref{4.17}) into
Eqs. (\ref{4.14}),(\ref{4.12}) we obtain the following microscopic
expressions for $\delta P$ and $\pi_{ij}$
\begin{eqnarray}
\delta P &=& (1 + F_{0})\sum_{\bf p}\frac{p^{2}}{3m^{*}}\delta n_{\bf p}^{L}
\equiv v_{s}^{2}\delta \rho,
\label{4.18} \\
\pi_{ij} &=& (1 + \frac{1}{5}F_{2})\sum_{\bf p}\left \{
\frac{p_{i}p_{j}}{m^{*}} - \frac{p^{2}}{3m^{*}}\delta_{ij}\right \}
\delta n_{\bf p}^{L},
\label{4.19} 
\end{eqnarray}
where $v_{s}^{2}= (1 + F_{0})p_{F}^{2}/3mm^{*}$ is the square of
velocity of sound \cite{Pines} and $\delta \rho = m\delta n$ is a
variation of the mass density. The equivalence in Eq. (\ref{4.18})
shows that the scalar part of the stress tensor $\delta P$ exactly
equals to the variation of the pressure of the Fermi liquid. 

Expressions for $\delta P$ and $\pi_{ij}$ (\ref{4.18}) and (\ref{4.19})
have the same form as for the gas case except the factors in 
front of the sums and the change of the bare mass to the mass of
a quasiparticle.
Thus the derivation of an equation for the stress tensor is 
straightforward. The final set of the generalized hydrodynamics equations
of a Fermi liquid in the second order in parameter $\gamma$ (\ref{2.17})
takes the form
\begin{eqnarray}
\partial _{t}\delta n &+&n_{0}\nabla_{k}v_{k}=0  
\label{4.20} \\
mn_{0}\partial_{t}v_{i} &+&\nabla_{i}\delta P+
\nabla_{j}\pi_{ij}-en_{0}\nabla_{i}\varphi =0  
\label{4.21} \\
\partial_{t}\delta P &+&K\nabla_{k}v_{k}=0  
\label{4.22} \\
\partial_{t}\pi_{ij} &+&\mu \left( \nabla_{i}v_{j}+
\nabla_{j}v_{i}-\frac{2}{3}\delta_{ij}\nabla_{k}v_{k}\right) =
-\nu_{c}\pi_{ij}.  
\label{4.23}
\end{eqnarray}
Hence, in a Fermi liquid we have again the system of equations which 
is similar to the Maxwell's theory of highly viscous fluids, but with the
bulk $K$ and the shear $\mu$ moduli directly related to microscopic 
Landau parameters
\begin{eqnarray}
K &=& mn_{0}v_{s}^{2}= \frac{5}{3}P_{0}\frac{1+F_{0}}{1+\frac{1}{3}F_{1}},
\label{4.24}\\
\mu &=& P_{0}\frac{1+\frac{1}{5}F_{2}}{1+\frac{1}{3}F_{1}},
\label{4.25}
\end{eqnarray}
where $P_{0}$ is a pressure of an ideal Fermi gas. 

The bulk modulus (\ref{4.24})
and the shear modulus (\ref{4.25}) exactly coincide with those of 
Ref.\ \onlinecite{Conti}, where they have been obtained by analyzing the
linear response function. In ideal Fermi gas
($F_{l}=0$) Eqs. (\ref{4.24}) and (\ref{4.25}) transform to the ideal gas
moduli (see Sec. III).

\section{Hydrodynamic theory of collective modes}

\subsection{Plasma waves in a homogeneous system}

The frequencies of collective plasma modes in a charged liquid have an 
order of magnitude of $\omega_{p}$ which commonly resides in the
high frequency (collisionless) region $\omega /\nu_c \gg 1$.
The plasma oscillations are solutions of 
the linearized system of the hydrodynamics equations. To obtain
the dispersion of plasmons we have to solve Eqs. (\ref{4.20})-(\ref{4.23})
with the collision term being omitted in Eq. (\ref{4.23}). In the
collisionless limit it is convenient to introduce a displacement
vector ${\bf u}$ and rewrite this system in the form of the 
elasticity theory (\ref{3.21}), (\ref{3.22}), where $K$ and $\mu$ are
determined, in general, by Eqs. (\ref{4.24}) and (\ref{4.25}).
Substituting Eq. (\ref{3.22}) to Eq.(\ref{3.21}) we have an equation
for the displacement vector of a "charged elastic medium"
\begin{equation}
mn_{0}\partial_{t}^{2}{\bf u}-(K+\frac{1}{3}\mu)\nabla (\nabla {\bf u}) -
\mu \nabla^{2}{\bf u} - en_{0}\nabla \varphi =0,
\label{5.1}
\end{equation}
which should be solved together with the Poisson equation
\begin{equation}
\nabla^{2} \varphi = -4\pi e \nabla n_{0}{\bf u}
\label{5.2}
\end{equation}
Considering a plane-wave solution $e^{-i(\omega t-{\bf qr})}$ we get 
the dispersion of the plasma waves
\begin{equation}
\omega^{2}(q)=\omega_{p}^{2}+v_{0}^{2}q^{2}
\label{5.3a}
\end{equation}
where
\begin{equation}
v_{0}^{2}=\frac{1}{mn_{0}}(K + \frac{4}{3}\mu)= v_{s}^{2} + 
\frac{4\mu}{3mn_{0}}.  
\label{5.3}
\end{equation}
The first contribution in $v_{0}^{2}$ 
equals to the square of the sound
velocity $v_{s}^{2}$ (the result of the ordinary hydrodynamics) 
and comes from the fluctuations of pressure $\delta P$ (see formulas
(\ref{4.18}) and (\ref{4.24})).
The second contribution arises from the traceless part of the 
stress tensor $\pi _{ij}$. 
The sum gives a correct coefficient which can be obtained from the
longitudinal response function of a charged Fermi liquid.\cite{Pines}
In the case of a noninteracting gas at arbitrary
degeneracy the expression for $v_{0}^{2}$ reduces to the formula 
$$
v_{0}^{2}=\frac{5P_{0}}{3mn_{0}}+\frac{4P_{0}}{3mn_{0}}=
\frac{3P_{0}}{mn_{0}}\equiv <v_{p}^{2}>.  
$$
which coincides with the result of the RPA (up to the second order of
$qu/\omega \sim \gamma$) and in a degenerate Fermi
gas gives the well known result $\frac{3}{5}v_{F}^{2}$. It is
straightforward to show that taking into account the third and fourth
moments in the linearized system of equations leads to the correct 
plasmon dispersion up to $q^{4}$.

\subsection{Plasma oscillations of a confined electron liquid}

In the case of one dimensional motion (e.g. for plasma waves in an
infinite medium which are considered in the previous subsection) 
the displacement 
vector ${\bf u}$ depends only on one coordinate. Hence, the equation
of motion takes the following simple form:
\begin{equation}
\partial_{t}^{2}{\bf u}- v_{0}^{2} \nabla^{2}{\bf u} - 
\frac{e}{m}\nabla \varphi =0.
\label{5.4}
\end{equation}
Eq. (\ref{5.4}) coincides with that of the linearized Bloch's theory. 
The only difference is that
the dispersion coefficient equals to the correct value $ v_{0}^{2}$
instead of a square of a velocity of sound $ v_{s}^{2}$.
It seems easy to phenomenologically improve the BHT. One could try to
replace the dispersion coefficient by the correct value $ v_{0}^{2}$
and hence obtain a correct dispersion of plasmon.  
\cite{Bennett,Eguiluz,Halperin,Dobson} However, from the general point of view
such an improvement cannot be consistent since the correct static
limit requires $ v_{s}^{2}$ as a coefficient in Eq. (\ref{5.4}).
The frequency-dependent coefficient was commonly used
in hydrodynamics calculations to recover both the
high- and the low-frequency regimes (see Ref.\ \onlinecite{Dobson} 
and references
therein). It has been then recognized \cite{Dobson} that the
theory with a frequency dependent dispersion coefficient violates
the harmonic potential theorem (HPT) in a spatially inhomogeneous case.
Nonetheless, it was believed that since the theory with the replaced coefficient
gives the correct plasmon dispersion in the infinite medium it should
hopefully give the proper description of the high-frequency plasma 
oscillations for
any geometries of a confined electron gas.\cite{Bennett,Eguiluz,Halperin}

As we show in the previous sections, the consistent theory which is
valid both in a high- and low-frequency regimes must inevitably include
the tensor fields of higher rank (e.g. $\pi_{ij}$ in the second order
theory). At low frequency $\pi_{ij}$ goes to zero that gives a
correct static limit. However,  in the high-frequency regime it does not
contain any small parameter and gives the contribution of the same
order of magnitude as the variation of the scalar pressure. Within the
linearized equations the contribution of the tensor $\pi_{ij}$ can
be interpreted as a shear stress of a Fermi system. Formally this
leads to more complicated structure of the differential operator
in Eq. (\ref{5.1}) than one of the operator expected from the BHT 
Eq. (\ref{5.4}).

In the present subsection we consider two 
simple examples of a collective motion of a Fermi liquid which is
confined in three and one dimensions. These examples show that the
collective eigenfrequencies
strongly differ from predictions of the standard hydrodynamics.   

Consider a homogeneous (in equilibrium state) electron liquid which is
confined by an external potential. The 
homogeneity of the equilibrium state can be reached for example with
the help of a positively charged background of the density equal to the 
density of
electrons $n_{0}$ or with the parabolic potential of a proper frequency
$\omega_{0}$. For example, in a three-dimensional spherically symmetric
case (parabolic quantum dot) the frequency 
$\omega_{0} = \omega_{p}/\sqrt {3}$. The two-dimensional
(parabolic quantum wire) and one-dimensional (quantum well) potentials
have frequencies $\omega_{0}= \omega_{p}/\sqrt {2}$ and
$\omega_{0}= \omega_{p}$ respectively.

Eigenmodes are solutions to equations 
(\ref{3.21}),(\ref{3.22}),(\ref{5.2}) with the proper boundary conditions.
Integration of Eqs. (\ref{3.21}) and (\ref{5.2}) across the surface
gives the set of boundary conditions 
\begin{eqnarray}
n_{j}\sigma_{ij}({\bf r}_{s})&=& 0,
\label{5.5}\\
{\bf n}\nabla \varphi^{>}({\bf r}_{s}) - 
{\bf n}\nabla \varphi^{<}({\bf r}_{s})&=&
4\pi en_{0}{\bf n}{\bf u}({\bf r}_{s}),
\label{5.6}\\
\varphi^{>}({\bf r}_{s})&=& \varphi^{<}({\bf r}_{s}),
\label{5.7}
\end{eqnarray}
where ${\bf r}_{s}$ is a coordinate of a surface point, $\varphi^{>}$
and $ \varphi^{<}$ are respectively the electrostatic potential 
$\varphi$ outside and inside the surface and ${\bf n}$ 
is the unit vector normal to the surface.
The boundary condition (\ref{5.5}) corresponds to the "free surface" and allows
an electron liquid to cross the boundary and oscillate with respect to the
equilibrium position. For example, the rigid oscillations of an electron
liquid ${\bf u}= {\bf u}_{0}e^{-i\omega t}$ (${\bf u}_{0}$ is a
constant vector) satisfy Eqs. (\ref{3.21}),(\ref{3.22}),(\ref{5.2}) and
conditions (\ref{5.5})-(\ref{5.7}). Such oscillations have the
frequency $\omega = \omega_{0}$. This fact apparently reflects the
HPT. 

The above system of equations and boundary conditions also
satisfy the HPT (or generalized Kohn theorem) in the most general
formulation, which was suggested by Dobson.\cite{Dobson} We
demonstrate this for a Fermi liquid confined by a spherically symmetric
harmonic potential with $\omega_{0}=\omega_{p}/\sqrt{3}$ and 
subjected to external homogeneous electric
${\bf E}$ and magnetic ${\bf H}$ fields. The HPT motion corresponds
to rigid displacement ${\bf u({\bf r},t)}= {\bf u}_{0}(t)$ of the
liquid. Vector ${\bf u}_{0}(t)$ is obviously the
center-of-mass coordinate. Since the stress tensor is equal to zero,
the boundary condition (\ref{5.5}) is trivially satisfied and equation
of motion (\ref{3.21}) has the form
\begin{equation}
m\partial^2_{t}{\bf u}_{0}-e{\bf H}\times \partial_{t}{\bf u}_{0} 
+e{\bf E}-e\nabla \varphi^{<}=0.
\label{5.1ad}
\end{equation}
The scalar potential $\varphi$ satisfies the Laplace equation and
electrostatic boundary conditions (\ref{5.6}), (\ref{5.7}). For rigid
motion these conditions coincide with the conditions on
the boundary of a homogeneously polarized sphere with a polarization
vector ${\bf P}=-en_{0}{\bf u}_{0}$. The soliton of this problem
inside the sphere is a depolarization field
$$
\nabla \varphi^{<}=\frac{4\pi}{3}{\bf P}=-\frac{4\pi en_{0}}{3}
{\bf u}_{0}.
$$
Substituting this solution into the Eq.(\ref{5.1ad}) we find that
the center-of-mass moves according to equation
\begin{equation}
m\partial^2_{t}{\bf u}_{0}-e{\bf H}\times \partial_{t}{\bf u}_{0} 
+e{\bf E}+\omega_{0}^{2}{\bf u}_{0}=0,
\label{5.2ad}
\end{equation}   
in exact correspondence with the HPT.\cite{Dobson}

To avoid a confusion we would like to outline that the boundary
condition (\ref{5.5}) assumes the parabolic potential outside
the region occupied by an electron liquid. Physically it can be realized,
for example, at the boundary of a depletion region in p-n junctions or
in artificial quantum structures. The condition (\ref{5.5}) is
not directly applicable to a model surface of metals with abrupt
change of the positive background at the boundary. The reason is a
jump of a potential at the surface which forbids electrons to penetrate
the boundary.  

Let us separate longitudinal and transverse components of the displacement
${\bf u}$
\begin{equation}
{\bf u} = \nabla \psi + {\bf u}_{t}, \qquad \nabla {\bf u}_{t}=0
\label{5.8}
\end{equation}
($\psi$ is a potential of the longitudinal component) and rewrite
the system of equations (\ref{3.21}),(\ref{3.22}),(\ref{5.2}) and
boundary conditions (\ref{5.5})-(\ref{5.7}) in terms of variables
$\psi$ and ${\bf u}_{t}$. We are looking for harmonic in time
solutions $\sim e^{-i\omega t}$.

In the region occupied by an electron liquid we have the system
of differential equations:
\begin{eqnarray}
\omega^{2}\psi + c_{l}^{2}\nabla^{2}\psi + \frac{e}{m}\varphi&=&0,
\label{5.9}\\
\omega^{2}{\bf u}_{t} + c_{t}^{2}\nabla^{2}{\bf u}_{t}&=&0,\quad 
\nabla {\bf u}_{t}=0,
\label{5.10}\\
\nabla^{2}(\varphi + 4\pi en_{0}\psi)&=&0.
\label{5.11}
\end{eqnarray}

Outside the surface the Laplace equation for the scalar potential
$\varphi$ must be satisfied
\begin{equation}
\nabla^{2}\varphi =0.
\label{5.12}
\end{equation}

All solutions of Eqs. (\ref{5.9})-(\ref{5.12}) must fulfill the
boundary conditions
\begin{eqnarray}\nonumber
(c_{l}^{2} - 2c_{t}^{2})&n_{i}&\nabla^{2}\psi + 
2c_{t}^{2}n_{j}\nabla_{j}\nabla_{i}\psi\\
&+&2c_{t}^{2}n_{j}(\nabla_{i}u_{t}^{j} + \nabla_{j}u_{t}^{i})=0,
\label{5.13}\\
{\bf n}\nabla (\varphi^{<} &+& 4\pi en_{0}\psi)=
{\bf n}\nabla \varphi^{>},
\label{5.14}\\
\varphi^{<}&=& \varphi^{>},
\label{5.15}
\end{eqnarray}
where all functions are taken at the surface. In Eqs. (\ref{5.9})-(\ref{5.15})
we introduced the notations:
\begin{eqnarray}
c_{l}^{2}&=&v_{0}^{2}=\frac{1}{mn_{0}}(K + \frac{4}{3}\mu),
\label{5.16}\\
c_{t}^{2}&=& \frac{\mu}{mn_{0}}.  
\label{5.17}
\end{eqnarray}

The boundary condition (\ref{5.13}) demonstrates a peculiarity of the
generalized hydrodynamics.
Longitudinal and transverse oscillations are, in general, mixed at
the surface. This effect is well known in the elasticity theory.
\cite{Landau} If the shear modulus is nonzero, the incident
purely longitudinal (transverse) wave transforms into the mixed
wave containing both longitudinal and transverse components.
As a result the structure of eigenmodes is changed in comparison
with the common hydrodynamics ($\mu =0$). The
qualitative structure of the wave is not changed only in the case
of a normal incidence (for example radial modes in the spherical symmetric
system which are purely longitudinal). However, even in this case 
the shear modulus contributes to the boundary condition.
The boundary condition for a purely longitudinal motion takes the
form
$$
(c_{l}^{2} - 2c_{t}^{2})n_{i}\nabla^{2}\psi + 
2c_{t}^{2}n_{j}\nabla_{j}\nabla_{i}\psi=0
$$  
which strongly differs from the common hydrodynamical condition
$\nabla^{2}\psi=0$ for a parabolic potential.\cite{Halperin}

Bellow we solve equations (\ref{5.9})-(\ref{5.15}) for two specific cases
and show that nonzero shear modulus strongly influences the dispersion
of the eigenmodes even for purely longitudinal waves.

\subsubsection{Radial plasma oscillations in a parabolic quantum dot.}

In a spherical symmetric quantum dot the normal vector 
${\bf n} = {\bf \hat r}$.
Thus all normal derivatives transform to the derivative with respect to
the radial coordinate $r$. The direction of the displacement 
${\bf u}$ for radial modes coincides with ${\bf \hat r}$. Hence
the transverse component ${\bf u}_{t}$ equals to zero and functions
$\psi$ and $\varphi$ depend only on $r$.

Radial solutions of Eqs. (\ref{5.9}), (\ref{5.11}), (\ref{5.12}) which are
regular at zero and at the infinity take the following general form
\begin{mathletters}
\label{5.18}
\begin{equation}
\psi(r) = A_{1} \frac{\sin qr}{r} + B_{1},\quad 
\varphi_{<}(r) = A_{2} \frac{\sin qr}{r} + B_{2},
\end{equation}
\begin{equation}
\varphi_{>}(r) = \frac{C}{r},
\end{equation}
\end{mathletters}
Substituting (\ref{5.18}) into Eqs. (\ref{5.9}), (\ref{5.11}) and into
boundary conditions for the scalar potential (\ref{5.14}), (\ref{5.15})
one gets the result
\begin{eqnarray}
\psi(r)&=& A_{1}\left \{ \frac{\sin qr}{r} - 
\frac{\omega_{p}^{2}}{\omega_{p}^{2}+c_{l}^{2}q^{2}}\frac{\sin qR}{R}
\right \}
\label{5.19}\\
\varphi_{<}(r)&=& A_{1}4\pi en_{0}\left \{ \frac{\sin qr}{r} - 
\frac{\sin qR}{R} \right \}
\label{5.20}
\end{eqnarray}
Here $R$ is the radius of the quantum dot.
The wave vector $q$ is related to the frequency $\omega$
\begin{equation}
\omega^{2}=\omega_{p}^{2}+c_{l}^{2}q^{2}.
\label{5.21}
\end{equation}
The last boundary condition (\ref{5.13}) in the spherical symmetric case
reads
$$
(c_{l}^{2} - 2c_{t}^{2})\nabla^{2}\psi(R) + 
2c_{t}^{2}\partial_{r}^{2}\psi(R)=0.
$$
This equation together with the solution (\ref{5.19}) gives the final
dispersion equation  
\begin{equation}
\tan qR = \frac{qR}{1 - (c_{l}/2c_{t})^{2}q^{2}R^{2}}
\label{5.22}
\end{equation}
which determine the allowed values of $q$ and, consequently, the
frequencies of the eigenmodes (\ref{5.21}). The corresponding result of
BHT can be obtained from Eq. (\ref{5.22}) in the limit 
$c_{t}\rightarrow 0$ and reads 
$$\tan qR=0.$$ 
The quantity $(c_{l}/2c_{t})^{2}$ which governs the difference of the 
last two dispersion equations can be expressed in terms of Landau 
parameters:
$$
(c_{l}/2c_{t})^{2} = \frac{1}{3}\left( 1 +\frac{5}{4}
\frac{1+F_{0}}{1+F_{2}/5}\right).
$$
This value is obviously far from infinity 
(as assumed in the BHT).
For example, in the case of a Fermi gas ($F_{l}=0$) we have
$(c_{l}/2c_{t})^{2}=3/4$ which is less then one.

\subsubsection{Surface plasma modes at the edge of a parabolic potential well.}

Let us consider the surface plasma oscillations at the edge of an
electron system confined by a one dimensional
parabolic potential with
$\omega_{0}=\omega_{p}$. The situation is common in
p-n junctions where the potential in the depletion region
is parabolic and density of electrons (holes) in n (p) regions
is approximately constant up to the corresponding boundary of the 
depletion region. Possible applications of the results to the
surface modes of the parabolic potential well is obvious. Within
BHT this problem has been considered in Ref.\ \onlinecite{Halperin}.

We suppose that the electron gas occupies the lower half space ($z<0$).
Let the surface wave propagate along $x$-axis 
$\psi,{\bf u}_{t}\sim e^{iqx}$. Due to the symmetry, the $y$-component
of the vector ${\bf u}_{t}$ equals to zero. Hence the solution of
Eq. (\ref{5.10}) can be taken as
\begin{equation}
{\bf u}_{t}={\bf a}e^{\kappa z}e^{iqx}
\label{5.23}
\end{equation}
where ${\bf a}=(a_{x},0,a_{z})$ is a two dimensional constant vector
and
\begin{equation}
\kappa^{2}= q^{2}-\omega^{2}/c_{t}^{2}.
\label{5.24}
\end{equation}
Due to the condition $\nabla {\bf u}_{t}=0$
(see Eq. (\ref{5.10})) the constants $a_{x}$ and $a_{z}$ are not independent:
$$
a_{x}=\frac{\kappa}{-iq}a_{z}.
$$ 

The general solution of Es.(\ref{5.9}, \ref{5.11}) takes the form
\begin{eqnarray}
\psi &=& \left(Ae^{qz} + Be^{pz} \right)e^{iqx},
\label{5.25}\\
\varphi_{<} &=& -\left(\frac{\omega^{2}m}{e}Ae^{qz} + 
4\pi en_{0}Be^{pz} \right)e^{iqx},
\label{5.26}
\end{eqnarray}
with
$$
p^{2}= q^{2} + (\omega_{p}^{2}-\omega^{2})/c_{l}^{2}.
$$
In the upper half-space we have the solution of the Poisson equation
(\ref{5.12}):
\begin{equation}
\varphi_{>}= Ce^{-qz}e^{iqx}. 
\label{5.27}
\end{equation}
The relationship of the constants $B$ and $C$ to the constant $A$ can
be easily obtained from the electrostatic boundary conditions 
(\ref{5.14}), (\ref{5.15}). The result for the potential of the
irrotational part of displacement $\psi$ is
\begin{equation}
\psi = A\left [ e^{qz} +
\left(1-\frac{\omega^{2}}{\omega_{s}^{2}} \right) e^{pz} \right ]e^{iqx}
\label{5.28}
\end{equation}
where $\omega_{s}=\omega_{p}/\sqrt{2}$ is the classical surface
plasmon frequency in the infinite wave length limit. 
Let us rewrite the boundary condition (\ref{5.13}) explicitly in the
case of the plane boundary :
\begin{eqnarray}\nonumber
(&c_{l}^{2}& - 2c_{t}^{2})n_{i}\nabla^{2}\psi + 
2c_{t}^{2}\partial_{z}^{2}\psi +4c_{t}^{2}\partial_{z}u_{t}^{z} =0,
\\ \nonumber
&c_{t}^{2}&(\partial_{x}\partial_{z}\psi +\partial_{x}u_{t}^{z}
+\partial_{z}u_{t}^{x})= 0.
\end{eqnarray}
Substitution of the solutions (\ref{5.23}) and (\ref{5.28}) to
the boundary conditions gives the system of equations for the
frequencies of the surface plasma modes:
\begin{eqnarray}
\left(1 - \frac{\omega^{2}}{\omega_{p}^{2}} \right)
\left(\omega^{2} -\omega_{s}^{2} -2c_{t}^{2}q^{2}\right)A -
 2c_{t}^{2}\kappa a_{z}=0,
\label{5.29}\\
c_{t}^{2}q^{2}
\left[q +\left(1 - \frac{\omega^{2}}{\omega_{s}^{2}} \right)p\right]A
+c_{t}^{2}\left(\kappa^{2} + q^{2}\right) a_{z}=0.
\label{5.30}
\end{eqnarray}
At $q=0$ we have
$$ 
\omega^{2} = \omega_{s}^{2}\equiv \frac{\omega_{p}^{2}}{2};\qquad a_{z}=0,
$$
which corresponds to the usual infinite wave length surface plasma
oscillations. At small $q$ the dispersion of the surface plasmon takes
the form
\begin{equation}
\omega^{2} = \omega_{s}^{2} + 2c_{t}^{2}q^{2}+
i\frac{4c_{t}^{3}q^{3}}{\omega_{s}}.
\label{5.31}
\end{equation}
The damping of the surface plasmon is not surprising since at $q\ne 0$
there is mixing with the transverse modes which can not decay from the surface
at high frequency. There are decaying transverse solutions only for
$\omega^{2}< c_{t}^{2}q^{2}$. Hence only the propagating transverse waves
are allowed at the frequency $\sim \omega_{s}$. The coupling of the
longitudinal and transverse components leads to the energy transfer
to  propagating waves and, consequently, provides damping of the
surface plasmon. It should be mentioned that the hydrodynamic theory
does not take into account the Landau damping due to the
single particle excitations. Inclusion of the Landau damping will
bring the usual contribution to the lifetime of the surface plasmon.
Yet in a weakly coupled electron gas the
Landau damping destroys the transverse modes. More precisely, if 
$c_{t}<v_{F}$ the transverse modes are damped. Hence the decay channel of
the surface plasmon into the transverse waves will be really open
if the inverse inequality $c_{t}>v_{F}$ is fulfilled and transverse
modes are well defined. Using Eq. (\ref{5.17}) for $c_{t}$ and
Eq. (\ref{4.25}) for $\mu$ we can express the condition $c_{t}>v_{F}$
in terms of Landau Fermi liquid parameters:
$$
F_{1}(1+F_{2}/5) + 3F_{2}/5 > 6.
$$
Consequently, the transverse modes and the decay channel described
above could exist for strongly correlated charged Fermi liquid. It
was mentioned in Ref.\ \onlinecite{Conti} that the condition 
should be satisfied for a electron liquid of low density.

With the accuracy up to $q^{2}$ we can neglect the imaginary term
in the dispersion law (\ref{5.31}). In this long wavelength approximation
the surface plasmon is a longitudinal wave with the dispersion
\begin{equation}
\omega^{2} = \omega_{s}^{2} + 2c_{t}^{2}q^{2}. 
\label{5.32}
\end{equation}
The last equation shows that the dispersion of the almost longitudinal
surface plasmon is totally determined by the shear modulus 
($c_{t}^{2}\sim \mu$). Within the BHT the shear modulus equals to zero.
In the case $c_{t}=0$ equations (\ref{5.29}, \ref{5.30}) lead to the
dispersionless  surface plasmon with the frequency $\omega = \omega_{s}$.
This agrees with the results obtained in \cite{Halperin} for
a wide parabolic quantum well. The correct dispersion law (\ref{5.32})
obtained from the generalized hydrodynamics has the dispersion
coefficient $2c_{t}^{2}$ which is of the same order of magnitude
as the coefficient $v_{0}^{2}=c_{l}^{2}$ in the volume plasmon
dispersion (\ref{5.3a}). For example, in the case of a high density
electron gas ($F_{l}\approx 0$) we have the ratio
$$   
2c_{t}^{2}/c_{l}^{2}= \frac{2}{3}.
$$ 
\section{The fourth-order theory. Thermal conductivity}

As we have seen in Sec. III, the theory of the second order 
in parameter $\gamma$ (\ref{2.17}) does not contain effects
of thermal conductivity which are responsible for the thermal 
equilibration processes. Hence the second-order approximation
cannot give correct static solution which should correspond
to a constant temperature ($\nabla T({\bf r})=0$).
To account for the heat conduction one has to consider at least
fourth-order approximation which contains two additional equations
for the third ${\bf L}^{(3)}$ and fourth ${\bf L}^{(4)}$ moments.
The system of equations follows Eqs. (\ref{2.13})-(\ref{2.16}).
The first two equations - the continuity equation and the equation
for the velocity field, are the same as in the second-order theory
(Eqs. (\ref{3.1}) and (\ref{3.2}) respectively). In the equations for
the scalar $P$ and traceless $\bbox {\pi}$ parts of the second
moment we get additional terms which are proportional to the gradient
of the ${\bf L}^{(3)}$ (see Eq. \ref{2.15}):
\end{multicols}
\widetext
\noindent
\raisebox{12pt}[0in][0.0in]
    {$\rule{3.4in}{0.4pt}\rule{0.4pt}{6pt}$\hspace{3.6in}}
\vspace{-0.15in}
\begin{eqnarray}
D_{t}P &+& \frac{5}{3}P\nabla\cdot {\bf v}+
\frac{2}{3}(\bbox {\pi} \cdot \nabla)\cdot {\bf v} +
Tr(\nabla \cdot {\bf L}^{(3)}) =0, 
\label{6.1}\\ 
D_{t}\bbox {\pi} &+& \bbox {\pi}(\nabla \cdot {\bf v}) +
\{(\bbox {\pi}\cdot \nabla)\otimes {\bf v}\}_{S} - 
{\bf 1}\frac{2}{3}(\bbox {\pi} \cdot \nabla)\cdot {\bf v}
+P\left( \{\nabla \otimes {\bf v}\}_{S}-
{\bf 1}\frac{2}{3}\nabla\cdot {\bf v} \right)+
\nabla \cdot {\bf L}^{(3)}-Tr(\nabla \cdot {\bf L}^{(3)}) = {\bf I}^{(2)}.
\label{6.2}
\end{eqnarray}
To close the set of equations with the accuracy of $\gamma^{4}$ we
should take an equation for ${\bf L}^{(3)}$ in the general form 
(\ref{2.16}) and omit the spatial derivatives of the fifth and the second
moments in the equation for ${\bf L}^{(4)}$:
\begin{eqnarray}
D_{t}{\bf L}^{(3)} &+& {\bf L}^{(3)}(\nabla \cdot {\bf v}) +
\{({\bf L}^{(3)}\cdot \nabla)\otimes {\bf v}\}_{S}
+ \nabla \cdot {\bf L}^{(4)}-
\frac{1}{mn}\{{\bf L}^{(2)}\otimes(\nabla \cdot {\bf L}^{(2)})\}_{S}  
= {\bf I}^{(3)},
\label{6.3}\\
D_{t}{\bf L}^{(4)} &+& {\bf L}^{(4)}(\nabla \cdot {\bf v}) +
\{({\bf L}^{(4)}\cdot \nabla)\otimes {\bf v}\}_{S} = {\bf I}^{(4)}.
\label{6.4}
\end{eqnarray} 
%
\begin{multicols}{2}
\noindent
Eqs. (\ref{3.1}), (\ref{3.2}) and (\ref{6.1})-(\ref{6.4}) constitute 
the system of the generalized hydrodynamics equations in the
fourth-order approximation.

The last term in the left-hand side of Eq. (\ref{6.1}) can be rewritten
as a divergency of a vector
\begin{equation}
Tr(\nabla \cdot {\bf L}^{(3)}) = \nabla_{j}L^{(3)}_{iij}=\nabla_{j}Q_{j},
\label{6.5}
\end{equation} 
where the vector $Q_{j}=L^{(3)}_{iij}$ describes the flow of energy
in a comoving frame.

To demonstrate the recovering of the thermal conductivity
we take the moments of the collision
integral ${\bf I}^{(3)}$ and ${\bf I}^{(4)}$ in a linear approximation,
similar to (\ref{3.7}):
\begin{eqnarray}
{\bf I}^{(3)} &=& -\nu_{c1}{\bf L}^{(3)},
\label{6.6}\\
{\bf I}^{(4)} &=& -\nu_{c2}\left ( {\bf L}^{(4)}-{\bf L}_{F}\right),
\label{6.7}
\end{eqnarray}
where we introduced the notation ${\bf L}_{F}$ for the fourth
moment of the distribution function $f_{\bf p}^{L}({\bf r},t)$ 
(see Eqs. (\ref{3.5})-(\ref{3.c})):
\begin{equation}
{\bf L}_{F}=\frac{1}{m^{3}}
\sum_{\bf p}p_{i}p_{j}p_{k}p_{l}f_{\bf p}^{L}
\label{6.8}
\end{equation}
Due to the spherical symmetry of the function $f_{\bf p}^{L}$ the fourth
rank tensor ${\bf L}_{F}$ (\ref{6.8}) takes the form:
\begin{eqnarray}
{\bf L}_{F}&=& 
(\delta_{ij}\delta_{kl}+\delta_{kj}\delta_{il}+\delta_{lj}\delta_{ik})L_{F},
\label{6.9}\\ \nonumber
\quad L_{F} &=& \frac{1}{15}\sum_{\bf p}\frac{p^{4}}{m^{3}}f_{\bf p}^{L}.
\end{eqnarray}
Parameters $\nu_{c1}$ and $\nu_{c2}$ in Eqs. (\ref{6.6})
and (\ref{6.7}) can be considered as phenomenological coefficients which
are, in general, not equal to the parameter $\nu_{c}$ (\ref{3.7}), but
have the same order of magnitude.

In the low frequency limit $\omega/\nu_{c}\ll 1$ we solve 
Eqs. (\ref{6.2})-(\ref{6.4}) using a perturbation theory. Since the
collision terms in all these equations have an order of magnitude of
the collision frequency the solution in zero order of $1/\nu_{c}$ 
follows from the equations
$$
{\bf I^{(j)}}=0,\quad j=2,3,4
$$ 
and takes the local equilibrium form
$$
{\bf L}^{(4)}={\bf L}_{F},\qquad {\bf L}^{(3)}=0,\qquad \bbox {\pi}=0.  
$$
The stress tensor becomes diagonal and proportional to the local pressure 
$$
{\bf L}^{(2)}={\bf 1}P_{F}
$$ 

The first order solution of equation (\ref{3.4}) for the 
tensor $\bbox {\pi}$
has been already obtained in Sec. III. It has the form of the
viscosity tensor (\ref{3.12}). Eq. (\ref{6.2}), which determines $\bbox {\pi}$
with the accuracy of $\gamma^{4}$, differs from Eq. (\ref{3.4}) 
since the third rank tensor ${\bf L}^{(3)}$ contributes in the 
left-hand side of (\ref{6.2}). However, the contribution of the third
moment in the Eq. (\ref{6.2}), being of the higher order in $\omega/\nu_{c}$,
does not change the first order result of Sec. III. 
Hence the viscous solution (\ref{3.12}) is still valid.

Let us consider solution of Eqs. (\ref{6.3}), (\ref{6.4}) for
the third moment ${\bf L}^{(3)}$.    
In the first-order of $\omega/\nu_{c}$ only the last two terms in the 
left-hand side in Eq. (\ref{6.3}) contribute to the equation
\begin{equation}
-\nu_{c1}{\bf L}^{(3)}\approx \nabla \cdot {\bf L}^{(4)}-
\frac{1}{mn}\{{\bf L}^{(2)}\otimes(\nabla \cdot {\bf L}^{(2)})\}_{S},
\label{6.10a}
\end{equation}
where zero-order expressions for ${\bf L}^{(4)}$ and ${\bf L}^{(2)}$
should be used in the right-hand side. 

Substituting zero-order (local equilibrium) solutions 
(${\bf L}^{(2)}={\bf 1}P_{F}$ and ${\bf L}^{(4)}={\bf L}_{F}$) to 
Eq. (\ref{6.10a}) one gets the 
the first-order expression for the third moment
\begin{equation} 
{\bf L}^{(3)} = -\frac{1}{\nu_{c1}}\left \{{\bf 1}\otimes
\left (\nabla L_{F} - \frac{1}{mn}P_{F}\nabla P_{F}  \right)\right \}_{S}.
\label{6.10}
\end{equation}
Contraction of the tensor ${\bf L}^{(3)}$ (\ref{6.10}) over the couple
of indexes gives the energy flow vector $Q_{j}=L^{(3)}_{iij}$,
which enters the equation of the  energy conservation (\ref{6.1})
\begin{equation} 
{\bf Q} = -\frac{5}{\nu_{c1}}
\left (\nabla L_{F} - \frac{1}{mn}P_{F}\nabla P_{F}  \right).
\label{6.11}
\end{equation}

Let us show that vector ${\bf Q}$ is proportional to the gradient
of temperature $\nabla T({\bf r},t)$ at arbitrary degeneracy of an
electron gas. The functions $L_{F}$ and $P_{F}$ depend on spatial
coordinates via the spatial dependence of the local chemical potential
$\mu({\bf r},t)$ and the local temperature $T({\bf r},t)$ entering
the distribution function $f_{\bf p}^{L}({\bf r},t)$
$$
f_{\bf p}^{L}({\bf r},t)=f^{L}\left(
\frac{\varepsilon_{\bf p}-\mu}{T}\right),
\qquad \varepsilon_{\bf p}=\frac{p^{2}}{2m}.
$$
Hence gradients of $P_{F}$ and $L_{F}$ can be rewritten as follows
\end{multicols}
\widetext
\noindent
\raisebox{12pt}[0in][0.0in]
    {$\rule{3.4in}{0.4pt}\rule{0.4pt}{6pt}$\hspace{3.6in}}
\vspace{-0.15in}
\begin{eqnarray}
\nabla P_{F}&=& \frac{1}{3m}\sum_{\bf p}p^{2}\nabla f_{\bf p}^{L}=
\left( -\nabla\mu + \frac{\mu}{T}\nabla T \right)\frac{1}{3m}
\sum_{\bf p}p^{2}\frac{\partial f_{\bf p}^{L}}{\partial \varepsilon_{\bf p}}
- \frac{\nabla T}{T}\frac{1}{6m^{2}}
\sum_{\bf p}p^{4}\frac{\partial f_{\bf p}^{L}}{\partial \varepsilon_{\bf p}}
\label{6.12}\\
\nabla L_{F}&=& \frac{1}{15m^{3}}\sum_{\bf p}p^{4}\nabla f_{\bf p}^{L}=
\left( -\nabla\mu + \frac{\mu}{T}\nabla T \right)\frac{1}{15m^{3}}
\sum_{\bf p}p^{2}\frac{\partial f_{\bf p}^{L}}{\partial \varepsilon_{\bf p}}
- \frac{\nabla T}{T}\frac{1}{30m^{4}}
\sum_{\bf p}p^{6}\frac{\partial f_{\bf p}^{L}}{\partial \varepsilon_{\bf p}}
\label{6.13}
\end{eqnarray}
\begin{multicols}{2} 
\noindent
The partial integration allows one to express the momentum
integrals in Eqs. (\ref{6.12}), (\ref{6.13}) in terms of macroscopic variables
$n$, $P_{F}$ and $L_{F}$:
\begin{eqnarray}\nonumber
\frac{1}{3m}\sum_{\bf p}p^{2}
\frac{\partial f_{\bf p}^{L}}{\partial \varepsilon_{\bf p}}&=&
-\sum_{\bf p}f_{\bf p}^{L}=-n,
\\\nonumber
\sum_{\bf p}p^{4}\frac{\partial f_{\bf p}^{L}}{\partial \varepsilon_{\bf p}}&=&
-5m\sum_{\bf p}p^{2}f_{\bf p}^{L}= -15m^{2}P_{F},
\\\nonumber
\frac{1}{30m^{4}}\sum_{\bf p}p^{6}
\frac{\partial f_{\bf p}^{L}}{\partial \varepsilon_{\bf p}}&=&
-\frac{7}{30m^{3}}\sum_{\bf p}p^{4}f_{\bf p}^{L}=-\frac{7}{2}L_{F}.
\end{eqnarray}
Using the last expressions we can represent Eqs. (\ref{6.12}) and (\ref{6.13})
in the following compact form
\begin{eqnarray}
\nabla P_{F}= \left( \nabla\mu - \frac{\mu}{T}\nabla T \right)n +
\frac{5P_{F}}{2T}\nabla T,
\label{6.13a}\\
\nabla L_{F}= \left( \nabla\mu - \frac{\mu}{T}\nabla T \right)\frac{P_{F}}{m} 
+ \frac{7L_{F}}{2T}\nabla T.
\label{6.14}
\end{eqnarray}
Comparison of Eqs. (\ref{6.13a}) and (\ref{6.14}) with 
Eq. (\ref{6.11}) shows that
the contribution of the gradient of the chemical potential exactly
cancels in the equation for the energy flow vector. Thus the vector
of the energy flow takes the usual form
\begin{equation}
{\bf Q} =-\lambda \nabla T
\label{6.15}
\end{equation}
with the coefficient of heat conduction
\begin{equation}
\lambda = \frac{5}{2\nu_{c1}T}\left(7L_{F} - \frac{5P_{F}^{2}}{mn}\right).
\label{6.16}
\end{equation}

Finally Eq. (\ref{6.1}) with $\bbox {\pi}$ (\ref{3.4}) and
${\bf Q}$ (\ref{6.16}) transforms to
the common hydrodynamical equation of the  energy conservation 
\cite{Landau}
\begin{equation}
D_{t}P + \frac{5}{3}P\nabla\cdot {\bf v}+
\frac{2}{3}(\bbox {\pi} \cdot \nabla)\cdot {\bf v} +
\nabla \cdot {\bf Q} =0,
\label{6.17}
\end{equation}
which contain both the viscous and the heat conduction terms.

The formula (\ref{6.16}) for the coefficient of heat conduction is
valid at arbitrary degree of degeneracy of an electron gas. 
In the nondegenerate case $P_{F}= nT$ and $L_{F}=nT^{2}/m$. 
Hence Eq. (\ref{6.16}) transforms to the result:
$$
\lambda = \frac{5nT}{\nu_{c1}m}.
$$
In a degenerate gas at $T=0$ we have $L_{F}= 5P_{F}^{2}/7mn$ hence
the expression in the brackets in (\ref{6.16}) cancels. At 
$T/\varepsilon_{F}\ll 1$ the correction has the order of magnitude
$(T/\varepsilon_{F})^{2}$ and the coefficient of 
heat conduction (\ref{6.16})
is proportional to $T/\nu_{c1}$. Since the collision frequency in a
degenerate Fermi system is a quadratic function of $T$ the heat
conduction diverges as $1/T$ (see, for example, Ref.\ \onlinecite{Pines}). 

Thus, in the low frequency limit the generalized hydrodynamics equations
(\ref{3.1}), (\ref{3.2}, (\ref{6.1})-(\ref{6.4})  coincides with the correct
set of the common hydrodynamics which consist of the continuity equation,
Navier-Stokes equation and the equation of the energy conservation in the
form (\ref{6.17}).

In the high frequency (collisionless) limit the generalized hydrodynamics
leads to the following  plasmon dispersion
$$
\omega^{2} = \omega_{p}^{2} + <v_{\bf p}^{2}>q^{2}
+\frac{<v_{\bf p}^{4}>-<v_{\bf p}^{2}>^{2}}{\omega_{p}^{2}}q^{4}
$$
that is exactly the result of the linear response theory.

The system of equations (\ref{3.1}), (\ref{3.2}), (\ref{6.1})-(\ref{6.4}) with
the collision terms (\ref{3.7}), (\ref{6.6}), (\ref{6.7}) constitute
the closed set of equations and provides the smooth interpolation
between high and low frequency regimes.

\section{Conclusion}
A description of dynamics of many-electron
systems in terms of macroscopic collective variables, which is usually
referred to as a hydrodynamical approach, provides a simple, physically
transparent and powerful tool for studying spatially inhomogeneous
problems. The common Bloch's hydrodynamics of
an electron gas is based on the assumption of a local equilibrium and
consequently shows a number of inconsistencies. We have shown
that it is possible to construct an inherently consistent generalized
hydrodynamics which correctly describes both the collisionless
high-frequency limit and the collision dominated
low-frequency regime, where the theory coincides with the 
standard Navier-Stokes
hydrodynamics. The theory follows from the long-wavelength expansion
of a kinetic equation and requires inclusion of new collective
variables with a nontrivial tensor structure. We remind that only 
scalar (pressure and density) and vector (velocity)
variables enter the common hydrodynamics. 

The appearance of higher
rank tensors is physically evident since the occupied region in a
momentum space loses its spherical symmetry under a general
(not locally equilibrium) evolution of the system. A need to describe
a non-spherical isoenergetic surface inevitably requires  tensor variables. 
As long as this surface is smooth it is possible to approximately 
describe it as a surface of a finite order and thus to take into
account the tensor fields of the finite rank. We have actually shown
above that the smoothness of the surface, which bounds the occupied
region in the momentum space, is governed by the basic parameter
(\ref{2.17}). 

In extension of the previous publication,\cite{hyd1} we presented a 
new general and more transparent derivation of
the generalized hydrodynamics and showed how the closed set of
equations of the
standard Navier-Stokes hydrodynamics (including the heat transport)
is recovered in the low frequency regime. 
We have shown that the generalized hydrodynamics can be built on the basis
of Landau theory of Fermi liquid which allows to determine
the correlation contribution to the stress tensor. Although
this hydrodynamics (as the Landau theory of Fermi liquid itself) is
meaningful only in the linear approximation, we belive that it could
be of a practical importance since it allows to express the results of 
hydrodynamical calculations, which are relatively simple, in terms of 
the microscopic Landau parameters. For example, the theory offers a
possibility to determine
the contribution of Fermi-liquid correlation effects to the eigenmodes of 
spatially inhomogeneous systems. 

In a linear approximation the tensor
variables, which describe the absence of a local equilibrium, can be
interpreted as an effective shear stress of a liquid. On a time scale
much longer than the collision time the contribution of the shear
stress vanishes and the dynamics of a liquid is governed only by the
usual bulk stress. However, as the collective modes of a charged
liquid normally
belong to the high frequency range, the shear contribution cannot be 
neglected. It is important to realize that this contribution cannot be
modeled, even qualitatively, merely by a change of the bulk 
modulus - the procedure, which has been
frequently used in literature to obtain the correct dispersion of
plasmon.\cite{Bennett,Eguiluz,Halperin} The reason is that the
relative contribution of the shear and bulk stresses is different for
different modes and different geometries. As a result the 
eigenfrequencies and the structure of plasma modes in
confined systems strongly deviate from predictions of BHT even with
``improved'' bulk modulus, which is adjusted to provide a correct plasmon
dispersion in a homogeneous situation. To demonstrate this 
qualitative non-local-equilibrium
effect we calculated eigenmodes of a Fermi liquid confined by a
harmonic potential of different dimensionality. The results show a
nontrivial contribution of the shear modulus to  eigenfrequencies of the
longitudinal plasma oscillation. The most transparent result concerns
surface waves at the edge of 1D harmonically trapped system. In the
absence of the shear modulus these modes are absolutely
dispersionless in agreement with the BHT results.\cite{Halperin} They
acquire 
a dispersion of the same order of magnitude as the bulk plasmon due to
deviation from the local equilibrium. We have also found that
existence of a nonzero shear modulus may open a new channel of decay
of these surface modes, in addition to the usual Landau damping.    
\\ \\  

The work of one of the authors (I.T.) was supported by the Alexander von
Humboldt Foundation.

\end{multicols} 
\end{document}